\documentclass[12pt,twoside,openright]{article}
\usepackage{feynmp}
\usepackage{epsf}
\usepackage{epsfig}
\usepackage{isolatin1}
\usepackage{amsmath}

\begin{document}

\title{Debye Screening mass from effective Potentials in Lattice gauge theory}
\author{\normalsize Dietrich Matthes\\ \normalsize Theoretische Physik III\\ \normalsize FAU Erlangen, Germany}
\maketitle
\begin{abstract}
In this paper, we use a definition of the Debye screening mass in $SU(2)$
pure gauge theory at high temperature in $3+1$ dimensions, that makes
use of the walls between different $Z_2$ domains. We show, that this
definition enables us to unambiguously obtain the Debye Screening mass from
the curvature of the effective potential for the Polyakov loops
calculated in the modified axial gauge. Using an approach on
asymmetric lattices, we obtain numerical values for the Debye mass and
compare them to perturbative as well as Monte-Carlo results. Our resulting
Debye-Masses are significantly higher than the one-loop perturbative
one. They are within the errors in agreement with those obtained
by Monte-Carlo methods. Open
questions are discussed.
\end{abstract}

\section{Introduction}
The Debye Screening mass in high Temperature $SU(2)$ pure gauge theory has
been subject of continued interest. The perturbative value in one
loop perturbation theory is known from calculations of the vacuum
polarization tensor \cite{shur} as well as other observables.\\
The perturbative definition of screening phenomena however remains unclear
with regard to infrared singularities in higher order perturbation theory
\cite{rebhan} as well as the convergence of the perturbation series in
non-abelian gauge theories in general.\\
Numerical investigations \cite{h2} \cite{h3} show, that there is a
significant discrepancy in the size of the Debye-Mass between calculations on a lattice using Monte-Carlo methods and the perturbative result. Up to
now, it remains unclear, to what extent higher order calculations may
overcome these differences or which non-perturbative effects might play a
role in the process of screening in non-abelian gauge theories even
at leading order.\\
It has been emphasized, that the Haar-Measure associated with the gauge
group SU(2) plays a major role when calculating properties of the
Polyakov loop in the modified axial gauge $\partial_0 A_0=0$ \cite{l3}.
This importance of the Haar measure stems from the fact, that in the
perturbative limit $g\rightarrow 0$ the non-abelian nature of the Gauge group
is inevitably lost, $SU(2)\rightarrow U(1)^{N-1}$, thus
effectively neglecting effects of the non-trivial Haar-measure. Since
it is this Haar-Measure, which leads to non-trivial topological effects
and thus to a feature, which is believed to be specific for a non-abelian Gauge
theory, it is desirable to take into account its effects also in the high-temperature phenomenon of Debye-Screening.\\
This paper is organized as follows. After the Introduction in part one, we
review a definition of the Debye-Mass via the profile of the domain
wall between different $Z_2$ domains of the Polyakov-Loop, which
fulfills the requirements of \cite{a3} and suits our approach well.
Furthermore, we show its consistency with perturbative calculations
in one loop perturbation theory. In the third part, we summarize some
results of \cite{b1} and \cite{hasen} on the method of asymmetric lattices,
which will be used further on. The fourth part contains our numerical
calculations of the curvature of the effective Potential for the
Polyakov-Loop and the results in terms of the physical temperature. The
fifth and last part is devoted to a summary of our results and a
discussion of open problems.

\section{Debye-Mass from the effective Potential of the Polyakov-Loops}

There are various suggestions for a definition of the Debye Screening Mass,
ranging from the exponential falloff of the correlator between two Polyakov
loops to the static limit of the longitudinal vacuum polarization tensor.
A requirement that has to be fulfilled by any of them to make sense is
its property of being odd under time reflection \cite{a3}. The definition,
that we want to use here has been suggested by \cite{b5} and
fulfils the abovementioned requirements. \\
It uses the fact, that the center symmetry is broken in the high temperature
phase in a $SU(2)$ gauge theory and the Polyakov loop may take on one of
the two possible $Z_2$-values. If there exist two different $Z_2$ phases
in space, then a domain wall $p(z)$ will develop between them. They can
be realized on a lattice by introducing a twist in the action and can thus
serve as a laboratory for measuring the Debye-Mass.\\
The Debye-Mass is then given by the exponential falloff of the profile of
the wall between different $Z_2$ domains, where here we will investigate only
domain walls developping in the $z$ direction $p(z)$
\begin{align}
m_D=\lim_{z\rightarrow\pm\infty}\left|\frac{\partial_z p}{2p}\right|.
\end{align}
The physics of this gauge-invariant non-perturbative definition is clear
in the abelian case. In the following, we will show in one loop perturbation
theory, that it coincides with the usual perturbative result in $SU(2)$. \\
This goal is achieved by calculating the explicit solution in a
Weiss-potential \cite{w1} extending in the $z$-direction and connecting two
different $Z_2$-domains. \\
Since in the modified axial gauge the vector potential $a_0\sim \tau^3$
we have to deal with a simple $(a_0^3)^4$-Theory, the field equation of which for a soliton is given by
\begin{align}
a_0''-\frac{\partial U}{\partial a_0} (z)=0
\end{align}
for the potential $U[a_0(z)]=\frac{a_0^4}{4}-\frac{m^2a_0^2}{2}+\mbox{const}$. The solution of this equation is
\begin{align}
&a_0(z)=\pm m\tanh\left[m(z-z_0)\right]. \label{vkrue2}
\end{align}
The potential energy of this solution is
\begin{align}
V[a_0]=\int dx\left(\frac{1}{2}(\nabla a_0)^2+U(a_0)\right).
\end{align}
Using a functional Taylor expansion for small fluctuations $\eta (z)$ around the minima $(a_0)_0$ of the potential
\begin{align}
V[a_0]=V[(a_0)_0]+\int dz\frac{1}{2}\left(\eta(z)\left[-\nabla^2+\left(\frac{d^2U}{d a_0^2}\right)_{(a_0(z))_0}\right]\eta(z)+...\right), \label{Taylor}
\end{align}
one can see immediately, that the exponential falloff $m$ of the solitonic
solution (\ref{vkrue2}) is given by the square root of the curvature of the potential at
the minimum
\begin{align}
m_D=\sqrt{\left.\frac{d^2 V_{eff}[a_0]}{d a_0^2}\right|_{\mbox{Min}}}.
\end{align}
Calculating this property for the effective potential in one-loop perturbation
theory \cite{w1}, one gets the result
\begin{align}
m_D=\sqrt{\frac{2}{3}}gT
\end{align}
which coincides with the one-loop result for the Debye-Mass as calculated with other definitions.\\
It is a peculiarity of the perturbative calculation, that the effect of the
nontrivial Haar-measure is cancelled exactly by the contribution of the
longitudinal gluons, thus leaving the range of integration in the
Path integral for the $a_0$ unclear. Studies of effective actions allowing for field
configurations with a nontrivial space-dependence \cite{m1}\cite{g2} (as compared to the
calculation of \cite{w1}, using $a_0(x)=const$.) have encountered
severe difficulties at these points $a_0=0,n\pi$. \\
In the following, we will therefore concentrate on the
non-perturbative lattice-calculation of an effective potential for the Polyakov loops, extracting
the Debye-mass from its curvature at the minimum subsequently.

\section{The method of asymmetric lattices}

Although huge efforts have been done to calculate an effective potential
for the Polyakov loops beyond perturbative methods, no satisfying success
has been reached yet except for the lattice approach. In this part, we
will use the calculation of the effective action on an asymmetric lattice
as described in \cite{b1}.\\
The essential idea of this method is the fact, that on a highly asymmetric
lattice with $N_t\ll N_s$ the lattice coupling $\beta$ is replaced by
two different spacelike and timelike couplings $\beta=\rho\beta_s=\frac{\beta_t}{\rho}$, where $\rho$ is the asymmetry parameter $\rho=\sqrt{\frac{\beta_t}{\beta_s}}$.
The quantum corrections \cite{k5} to these equations will not be
considered here. Using the abovementioned equations and the fact, that on
a highly asymmetric lattice $\beta_t \gg\beta_s$, a perturbative series of
the action
\begin{align}
S_W=\sum_{\vec{x},i}\left(\sum_i\beta_t\frac{1}{N}Tr G_{0i}(\vec{x},t)+\sum_{i<j}\beta_s\frac{1}{N}Tr G_{ij}(\vec{x},t)\right). \label{W}
\end{align}
in the spacelike coupling $\beta_s$ can be performed. Only considering the
order $O(\beta_s^0)$ for the moment, the spacelike link-variables $U_i(\vec{x},t)$ of the field strengths originating at $\vec{x}$ in direction $i,j$ respectively 
\begin{align}
&G_{0i}(\vec{x},t)=V(\vec{x},t)U_i(\vec{x},t+1)V^{\dagger}(\vec{x}+i,t)U_i^{\dagger}(\vec{x},t), \nonumber\\
&G_{ij}(\vec{x},t)=U_i(\vec{x},t)U_j(\vec{x}+i,t+1)U_i^{\dagger}(\vec{x}+j,t)U_j^{\dagger}(\vec{x},t),
\end{align}
can be integrated. One thus obtains \cite{b1}
\begin{align}
\exp (S_{eff}^0)=&\int\prod_{x,i;t}\left[DU_{x,i;y}\left(1+\sum_{j=\frac{1}{2}}^{\infty}d_j\frac{I_{2j+1}(\beta_t)}{I_1(\beta_t)}\chi_j(U_{x,i;t}V_{x+i;t}U^{\dag}_{x,t+1;i}V^{\dag}_{x,t})\right)\right]= \nonumber\\
=&\prod_{x,i}\left(1+\sum_{j=\frac{1}{2}}^{\infty}
\left[\frac{I_{2j+1}(\beta_t)}{I_1(\beta_t)}\right]^{N_t}\chi_j(P_{x+i})\chi_j(P_x^{\dag})\right)
\end{align}
with $\chi_j$ the characters in $j$th representation of the Polyakov loops
\begin{align}
P_x=\left(\begin{array}{cc} e^{i\theta_x} & 0 \\ 0 & e^{-i\theta_x} \\ \end{array}\right).
\end{align}
This result, first obtained by \cite{b1} cannot be simplified analytically
for general $N_t$. \\
This is why we used $\theta_x=\theta_{x+i}$ to calculate the effective potential
\begin{align}
V_{eff}=-\frac{1}{V}\ln\left(\left(\prod_x \sin^2(\theta)\right)\exp(S_{eff}^0)\right) \label{v}
\end{align}
for general $N_t$. \footnote{A mean field analysis as performed
  in \cite{b1} for the critical coupling yields the same qualitative phase structure but is less practical for calculating the curvature of the potential
  at the minimum.} The $O(\beta_s^0)$ also coincides with the result given
in \cite{g3} for a strong coupling approximation. The reason for this to be
the case is the fact that the $O(\beta_s^0)$ effectively amounts to neglecting
the magnetic parts of the field strength. This is exactly how the result \cite{g3} has been obtained. 
\begin{figure}[h]
\begin{minipage}[t][11cm][t]{0.6cm}
  \vspace{4cm}
$V_{eff}$
\end{minipage}
\begin{minipage}[t][11cm][t]{12cm}
\epsfig{file=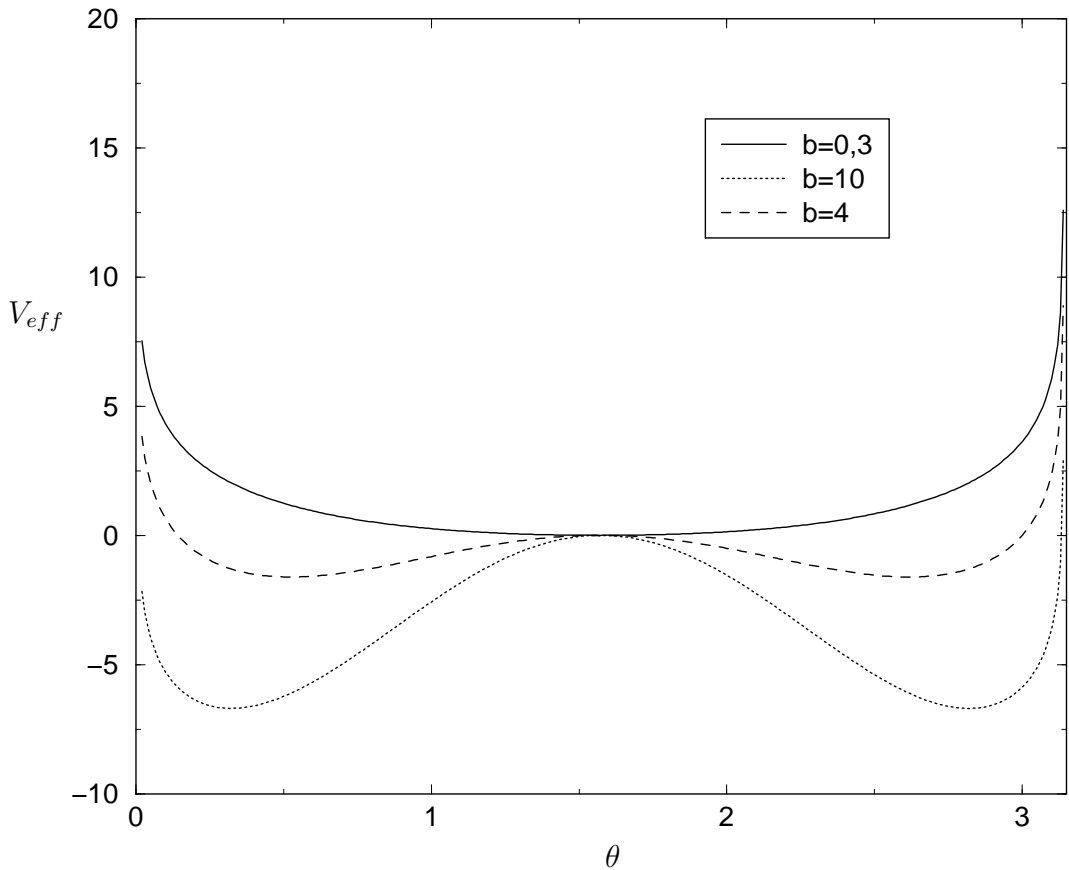,width=11cm,angle=-90}
\end{minipage}
\hfill
\begin{minipage}[t][11cm][t]{0.1cm}
\end{minipage}
\begin{minipage}[b][0.7cm][c]{14cm}
\hspace{7.8cm}
$\theta$
\end{minipage}
\parbox{15cm}{\caption{\label{vstrong}Potential $V_{eff}$ in
    approximation $O(\beta_s^0)$
    for different couplings b with a timelike extension $N_t$ of
    the lattice. For increasing coupling b (increasing temperature), the potential shows a second order phase transition to a phase, where the center
    symmetry is spontaneously broken.}}
\end{figure}

\section{Numerical results}

We now proceed investigating the properties of the effective potential
numerically. We obtain the effective potential for the
Polyakov Loops numerically by only considering spatially constant
field configurations $\theta_x=\theta_{x+i}$.\\
Then we search for the minima of the effective
potential as a function of $\beta_t$ for different
values of $N_t$. After that, we calculate the curvature at the minimum by
twice taking the derivative of $V_{eff}$ with respect to the argument $\theta$ at the minimum. Next, we relate the lattice coupling to the physical
temperature $\frac{T}{T_c}$ and fit the data to the functional form of the
leading order (one loop) perturbative result for the Debye Mass
\begin{align}
m_D=C\sqrt{\frac{2}{3}}g(T)T. \label{mdpt}
\end{align}
This procedure deserves a more detailed
explanation. \\
We
are eventually interested in the prefactor $C$. Previous lattice calculations \cite{h2}, \cite{h3}
suggested a $C\in [1.6,2]$ as compared to the one loop perturbative result $C=1$.\\
For being able to make reliable predictions about the size of this
prefactor, one has to compare results at the same physical temperature. We
follow here an approach outlined in \cite{e1}. \\
Violations of asymptotic
scaling are taken into account by considering the general Ansatz
\begin{align}
&a\Lambda=R(g^2)\lambda(g^2) \label{alambda}\\
&R(g^2)=\exp\left(\frac{-b_1}{2b_0^2}\ln(b_0g^2)-\frac{1}{2b_0g^2}\right), \label{R}\\
&b_0=\frac{11N_c}{48\pi^2}, \;\; b_1=\frac{34}{3}\left(\frac{N_c}{16\pi^2}\right)
\end{align}
The function $\lambda(g_t^2)$ parametrizes the asymptotic scaling violations,
the values for $b_0$ and $b_1$ are found perturbatively. For $\lambda(g_t^2)$, we use an exponential ansatz
\begin{align}
\lambda(g_t^2)=\exp\left(\frac{1}{2b_0^2}(d_1g_t^2+d_2g_t^4+d_3g_t^6)\right).
\label{lambda}
\end{align}
With $T=\frac{1}{N_ta_t}$, we obtain
\begin{align}
\frac{1}{N_t R(g_{t,c}^2)}=\lambda(g_{t,c}^2)\frac{T_c}{\Lambda}.
\end{align}
The value of the coupling $g_{t,c}^2$ at the critical Temperature $T_c$ of the
deconfinement phase transition can be obtained numerically by inspection
of the effective potentials for various $N_t$ (see figure \ref{vstrong}).
We thus arrive at the fit parameters $d_1= 0.0068,\, d_2= -0.001173,\, d_3= 6.4818 \, \, 10^{-5}$ in equation (\ref{lambda}). \\
Now we can extract the temperature in units of the critical temperature for
a given $N_t$. It is given by
\begin{align}
\frac{T}{T_c}=\frac{R(g_{t,c}^2)\lambda(g_{t,c}^2)}{R(g_t^2)\lambda(g_t^2)}. \label{tdurchtc}
\end{align}
This sets us in the position to relate the Temperature $T$ to the
coupling $\beta_t$. This temperature scale can now be used to relate the lattice Debye mass
given by the curvature of the effective potential (\ref{v}) at the minimum in
terms of the asymmetric lattice coupling $\beta_t$ to the physical
temperature $T$. \\
We now use the functional dependence of the Debye-mass on $g(T)$ and $T$ as
known from one-loop perturbation theory to determine $C$ in equation (\ref{mdpt}). To this end, we need the coupling as a function of the temperature $g(T)$.
For this running coupling, we use the two-loop formula
\begin{align}
 g(T) = \left\{\frac{11}{12\pi^2}\left(\ln\left(2\pi\frac{T}{T_c}\frac{T_c}{\Lambda}\right)\right)+\frac{17}{44\pi^2}
   \left(\ln\left[2\ln\left(2\pi\frac{T}{T_c}\frac{T_c}{\Lambda}\right)\right]\right)\right\}^{-\frac{1}{2}}. \label{gvont}
\end{align}

\begin{figure}[h]
\begin{minipage}[t][10cm][t]{0.6cm}
  \vspace{4cm}
$C$
\end{minipage}
\begin{minipage}[t][10cm]{11.7cm}
\epsfig{file=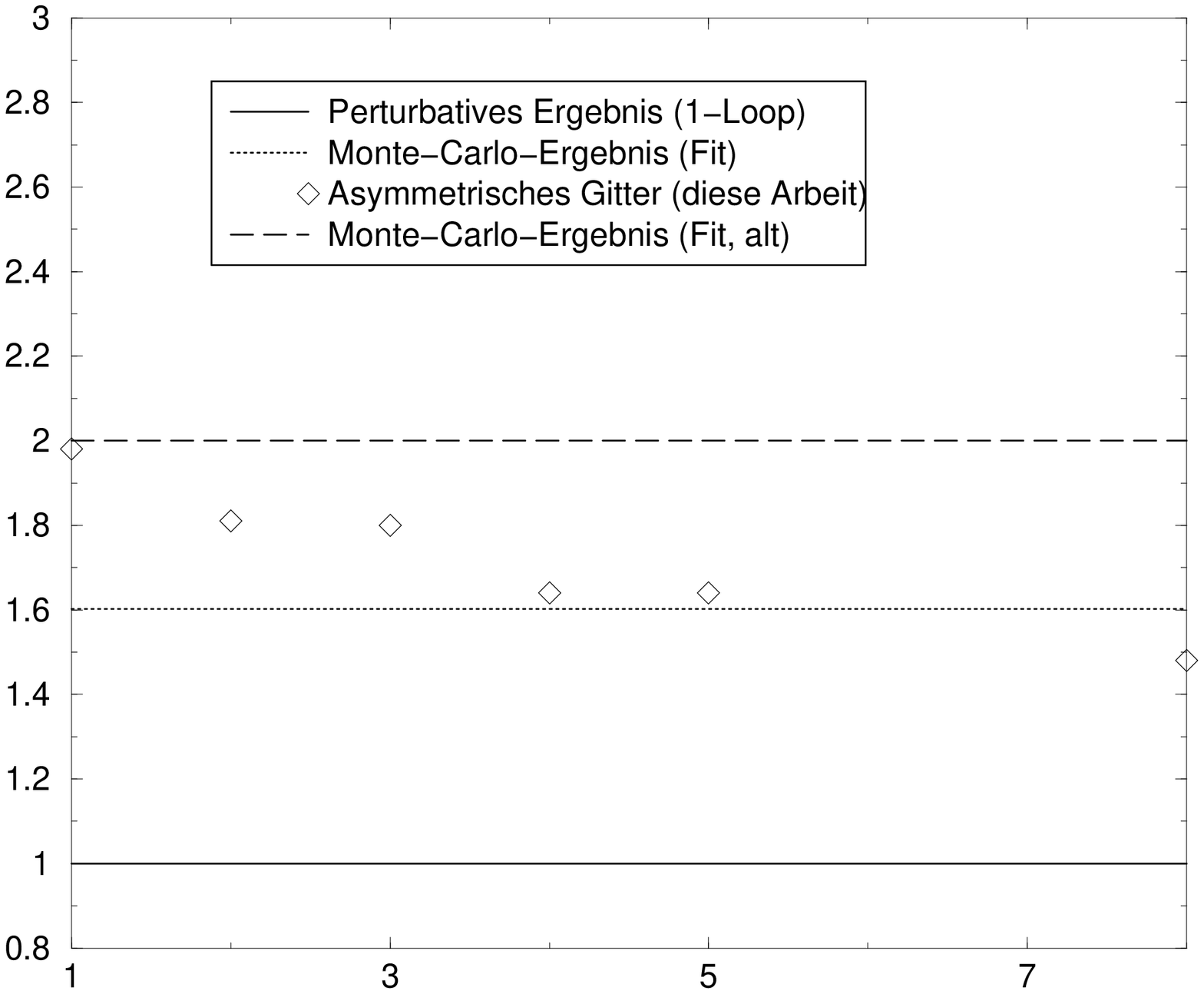,width=10cm,angle=-90}
\end{minipage}
\hfill
\begin{minipage}[t][10cm][t]{0.011cm}
\end{minipage}
\begin{minipage}[b][0.7cm][c]{14cm}
\hspace{7.1cm}
$N_t$
\end{minipage}
\parbox{15cm}{\caption{\label{mdvonntbild2}Comparison of the Debye masses
    obtained for various values of $N_t$. The Value of \cite{h2} and the
    older one of \cite{h3} as
    well as the perturbative value are indicated, too.}}
\end{figure}

\begin{center}
\begin{table}
  \begin{center}
\begin{tabular}[t]{|c|c|c|c|}
\hline
$N_t$ & $\beta_c$ & $\frac{m_D}{m_D^{PT}}$ & $\frac{T_c}{\Lambda_{\overline{MS}}}$ \\ \hline \hline
 1 &  0,39  &  1,98 & 1,013 \\ \hline
 2 &  1,29  &  1,81 & 0,889 \\ \hline
 3 &  2,09  &  1,80 & 0,983 \\ \hline
 4 &  2,54  &  1,64 & 1,156 \\ \hline
 5 &  3,18  &  1,64 & 1,330 \\ \hline
 8 &  5,37  &  1,48 & 1,556 \\ \hline
\end{tabular}
\end{center}
\begin{center}
\parbox{5cm}{\caption{\label{tabelle}Results}}
\end{center}
  \end{table}
\end{center}
Using (\ref{alambda}), (\ref{tdurchtc}) and (\ref{gvont}) in (\ref{mdpt}),
we can determine the prefactor $C$ which gives the quotient of $m_D$ with
the perturbative value $m_D^{p.t.}$. \\
Carrying out this procedure for several values of $N_t$ in the temperature
range $2<\frac{T}{T_c}<8$, 
we arrive at the values for the prefactor $C$ in the one loop perturbative
Debye mass given in table \ref{tabelle}. These results are shown
in figure \ref{mdvonntbild2}. The values for $\frac{T_c}{\Lambda_{\overline{MS}}}$, that we
obtained with our fit are also indicated in the table.
A comparison with the numerical result $\frac{\Lambda_{\overline{MS}}}{\Lambda_{W}}=19.8231 $ \cite{lambda} has been performed. Thus the deviation from
one gives the degree of accuracy and shows, that it agrees well with the expection for small
lattices, whereas it strongly disagrees for larger lattices. This will
also be adressed in the next section.

\section{Discussion and Outlook}

\begin{figure}[h]
\begin{minipage}[t][11cm][t]{0.6cm}
\vspace{4.8cm}
$\frac{V_{eff}^{(0)''}}{V_{eff}^{(2)''}}$
\end{minipage}
\begin{minipage}[t][11cm][t]{12cm}
\epsfig{file=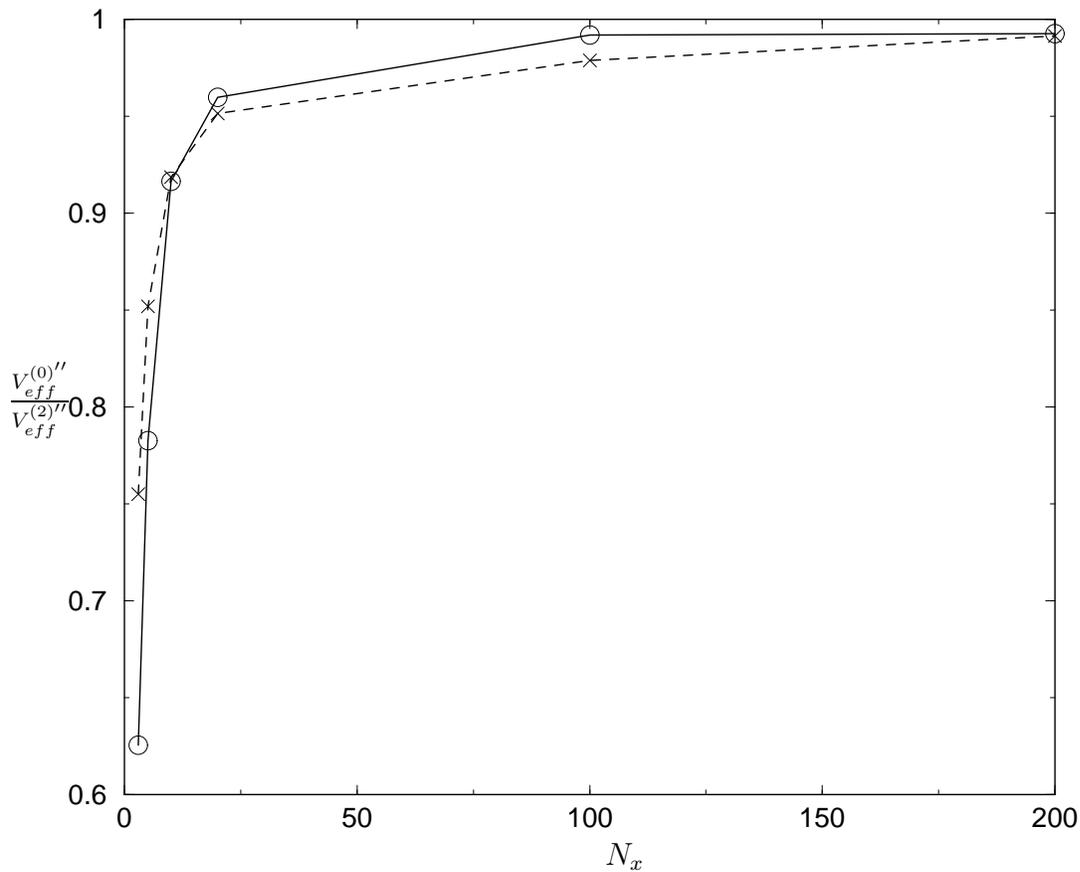,width=11cm,angle=-90}
\end{minipage}
\hfill
\begin{minipage}[t][11cm][t]{0.1cm}
\end{minipage}
\begin{minipage}[b][0.7cm][c]{14cm}
\hspace{7.8cm}
$N_x$
\end{minipage}
\parbox{15cm}{\caption{\label{veffobetaquadr}Difference of the curvature
    of the effective potentials in $O(\beta_s^0)$, $V_{eff}^{(0)''}$, and
    $O(\beta_s^2)$,  $V_{eff}^{(2)''}$, at the minimum for $N_t=5$ (full line) and $N_t=3$ (dashed line), $\frac{\beta}{\beta_c}=11,5$. For $\rho =\frac{N_x}{N_t}=1$, the deviation is  22\% at $N_t=5$ and 24\% at $N_t=3$, for $\rho =20$ only 0,81\% and 3,8\% respectively.}}
\end{figure}

Using a definition of the Debye-Mass which allows to extract it from
the gauge invariant effective potential for the Polyakov loops, we
showed by explicit solution, that the perturbative result is reproduced, when the Weiss-Potential is considered. \\
We then used an asymmetric lattice to determine the effective potential nonperturbatively. To this end, we only considered the leading order
contribution in an expansion in the space-like lattice coupling. This
approximation is believed to be of good accuracy for highly asymmetric
lattices. This can also be shown numerically by inspection of the
curvatures of the corresponding effective potentials in $O(\beta_s^0)$ and
$O(\beta_s^2)$ for different values of the asymmetry parameter $\rho$ (which
ultimately give the Debye mass).
The result is depicted in figure \ref{veffobetaquadr}. The influence
of the next to leading order on the temperature scale is not taken into
account here.\\
Our investigation then proceeded by numerical calculation of the
Debye-Mass. 
The results of our calculation yield a prefactor which is significantly
larger than the perturbative result. It is within the error bars due to
the fits, that we performed compatible with the 
result suggested by full Monte-Carlo calculations \cite{h2}, which is
also shown in figure \ref{mdvonntbild2}. \\
The major shortcoming of the method used here is the fact, that
no Monte-Carlo values of $\frac{T_c}{\Lambda}$ are available for highly 
asymmetric lattices (only these can be hoped to give reliable predictions 
for the Debye-Mass). Thus there is no control as to whether the temperature
scale is of good accuracy and whether the results for
$C$ are reliable. \\
In addition to that, since we used the $O(\beta_s^0)$ of the asymmetric
expansion only, we are effectively working in the strong coupling regime and
neglecting magnetic terms in the action. Due to this, our results are
not obtained in the scaling regime, which can also be seen from the
funtion $\lambda (g_t^2)$ (\ref{lambda}), which fits the asymptotic scaling violation and is
not constant. This behaviour of the function parametrizing the
asymptotic scaling violation is also seen in full numerical
investigations \cite{h2}\cite{h3}. However, this problem can easily be overcome by
incorporating the contribution of the $O(\beta_s^2)$ terms. Our
investigation of the errors of the truncation of the series in $\beta_s$
shows, that the effect on the Debye Mass will be small on the scale
of the discrepancy between perturbative continuum result and full
lattice Monte-Carlo result. \\
Furthermore, we had to do several fits, which are of varying accuracy. In this
particular case, we performed the analysis for rather small values of
$\beta_t$, because the expansion of the action in $\beta_s$ is better for
small $\beta$. In addition to that, we used perturbative results for
$R(g^2)$ as well as for $g(T)$. It is unclear, whether the coupling range that
we investigated is in the perturbative regime already.\\
On the other hand, the results show, that there is a strong influence of
the wall at PL$\approx\pm1$ in the effective Potential for the PL associated to the non-trivial Haar-measure on the size of the
inverse screening length. Since the fact, that this wall does not vanish is
a major property of non-perturbative treatments of the action (as has
also been seen in a path-integral approach in $1+1$ dimensions \cite{martin} and in a new, remarkable calculation \cite{Kenji}), the
difference in size of the Debye-Mass can be attributed to this Haar-measure,
which inevitably modifies the curvature at the minimum of the effective
potential.
\subsubsection*{Aknowledgements}
I would like to thank the Studienstiftung des Deutschen Volkes for
a PhD research grant.

\end{document}